\documentstyle[12pt,epsf,epsfig,wrapfig]{article}
\begin{document}

\newcommand{\tv}[1]{{\bf #1}_{T}}
\newcommand{\nnn}{\nonumber\\}
\newcommand{\Pslash}{\kern 0.2 em P\kern -0.56em \raisebox{0.3ex}{/}}
\newcommand{\kslash}{\kern 0.2 em k\kern -0.45em /}
\newcommand{\Sslash}{\kern 0.2 em S\kern -0.56em \raisebox{0.3ex}{/}}

\vspace*{-15mm}
\hspace*{\fill} NIKHEF 96-025\\
\hspace*{\fill} hep-ph/9610295\\[-7mm]

\begin{center}
{\large \bf
Time reversal odd fragmentation functions
\footnote{Contributed paper at the 12$^{th}$ International Symposium
on High Energy Spin Physics, Amsterdam, Sept. 10-14, 1996}
\\ }
\vspace{5mm}
\underline{R. Jakob}$^1$ and P.J. Mulders$^{1,2}$
\\
\vspace{5mm}
{\small\it
(1) NIKHEF, P.O.Box 41882, 1009 DB Amsterdam, The Netherlands\\
(2) Free University, De Boelelaan 1081, 1081 HV Amsterdam, The Netherlands
\\ }
\end{center}

\begin{center}
ABSTRACT

\vspace{5mm}
\begin{minipage}{130 mm}
\small
The combination of transverse momentum and polarization effects in hard
scattering processes, e.g polarized deep-inelastic leptoproduction, results in
a rich variety of information on the hadronic structure. This information is
encoded in a correspondingly large number of functions depending
on both, longitudinal fractional momenta $x$ and transverse momenta
$p_T$. Integration over transverse momenta establishes the connection to
the usual distribution and fragmentation functions. Constraints from
hermiticity and invariance under parity operation induce some relations between
different functions.\\  
One specific aspect of the information on the hadronic structure is the
appearance of {\em time reversal odd} fragmentation functions due to the
non-applicability of time-reversal invariance for the hadronization of a
quark. {\em T-odd} fragmentation functions are experimentally accessible
via the measurement of, for instance, asymmetries in semi-inclusive
leptoproduction. 
\end{minipage}
\end{center}

\begin{wrapfigure}{r}{6cm}
\centerline{
\epsfig{figure=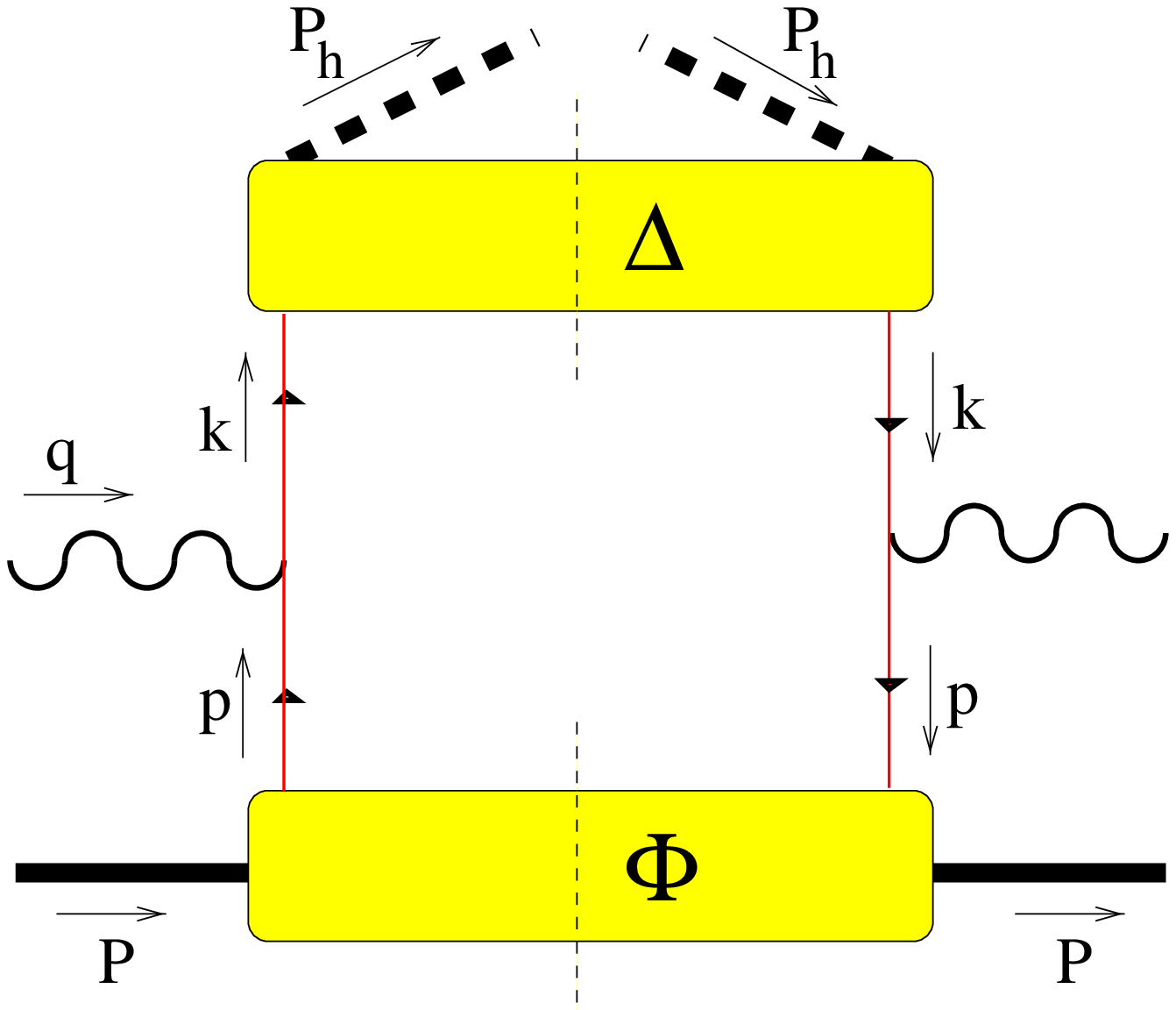,width=4.5cm}}
{\small Figure 1: Diagram of the leading conributions to 
1-particle inclusive leptoproduction.}
\end{wrapfigure}

The basic ingredient of the diagrammatic approach to hard scattering processes
[1] is the assumption that the hadronic tensor {\em factorizes} in {\em hard}
and {\em soft} parts separated by different powers of a hard scale $Q$. The
information on the hadronic structure is encoded in hadronic matrix elements
of quark (and, in general, gluon) fields, conveniently parametrized in
distribution (DF) and fragmentation functions (FF). The leading contributions to
1-particle inclusive leptoproduction from the 'handbag' diagram (Fig.1)
involve the non-local hadronic matrix elements
\begin{equation}
\Phi_{ij}(p,P,S)=\frac{1}{(2\pi)^4}\int d^4x \;
e^{ip\cdot x}\;\langle P,S|\bar\psi_j(0)
\psi_i(x)|P,S\rangle
\end{equation}
for the distribution of quarks in the target hadron (with momentum $P$ and 
spin vector $S$) 
and
\begin{equation} 
\Delta_{ij}(k,P_h,S_h)=\sum_X\frac{1}{(2\pi)^4}\int d^4x\;
e^{ik\cdot x}\;\langle 0|
\psi_i(x)|P_h,S_h;X\rangle
\langle P_h,S_h;X|\bar\psi_j(0)|0\rangle
\end{equation}
describing the decay of a quark, thereby, amongst others producing the one
observed hadron in the final state (characterized by $P_h$ and $S_h$). Both 
objects, $\Phi$ and $\Delta$, are
hermitian and reveal symmetries under discrete transformations: both are parity
invariant, whereas only $\Phi$ is invariant under {\em time
reversal}, as well. The {\em time reversal} operation relates {\em in}-states
to {\em out}-states and thus $\Delta$ has no simple symmetry properties under
this operation. The most general ansatz for the quark-quark correlation function
$\Delta(k;P_h,S_h)$ consistent with the hermiticity properties of the field
and invariance under parity operation is
($k$ is the momentum of the quark) [2]
\begin{eqnarray}
\lefteqn{
\Delta(k;P_h,S_h)=
B_1 M_h + B_2 \Pslash_h +B_3 \kslash 
+ (B_4/M_h)\; \sigma_{\mu\nu}P_h^\mu k^\nu} 
\nnn &&
{}+ i\; B_5 (k\cdot S_h)\gamma_5
+ B_6 M_h \Sslash_h \gamma_5 
+ (B_7/M_h)(k\cdot S_h)\Pslash_h \gamma5
+ (B_8/M_h)(k\cdot S_h)\kslash \gamma_5 
\nnn &&
{}+ i\; B_9\;\sigma_{\mu\nu}\gamma_5 S_h^\mu P_h^\nu
+ i\; B_{10}\; \sigma_{\mu\nu}\gamma_5 S_h^\mu k^\nu 
+ i\; (B_{11}/M_h^2)(k\cdot S_h)\; \sigma_{\mu\nu}\gamma_5 k^\mu P_h^\nu
\nnn &&
{}+(B_{12}/M_h)\; \epsilon_{\mu\nu\rho\sigma} 
\gamma^\mu P_h^\nu k^\rho S_h^\sigma.
\end{eqnarray}
Hermiticity requires all the coefficient functions $B_i$ to be real. The
additional constraint of time reversal invariance, if applicable, would 
imply ${B_i}^\ast=-B_i$
for $B_4$, $B_5$, and $B_{12}$. In this sense, those coefficient functions are
refered to as {\em time reversal odd} (or, in short, {\em T-odd}).\\
The object appearing in the diagrammatic expansion is the $k^+$-integrated
hadronic matrix element, conventionally, parametrized in 
terms of generalized FF's (up to ${\cal O}(1/Q)$)
\begin{eqnarray}
\lefteqn{ 
\frac{1}{4z}\int dk^+\Delta(k,P_h,S_h)\Bigg|_{k^-=\frac{P_h^-}{z},\tv{k}}=
\frac{M_h}{4P_h^-}\Bigg\{
E+D_1\frac{\Pslash_h}{M_h}
+D_{1T}^\perp\frac{\epsilon_{\mu\nu\rho\sigma}   
\gamma^\mu P_h^\nu k_T^\rho S_{hT}^\sigma}{M_h^2}}
\nnn &&
{}+D^\perp\frac{\kslash_T}{M_h}
+D_T\epsilon_{\mu\nu\rho\sigma}                  
n_+^\mu n_-^\nu\gamma^\rho S_{hT}^\sigma
+\lambda_h D_L^\perp\frac{\epsilon_{\mu\nu\rho\sigma}   
n_+^\mu n_-^\nu \gamma^\rho k_T^\sigma}{M_h}
-E_si\gamma_5                                    
-G_{1s}\frac{\Pslash_h\gamma_5}{M_h}
\nnn &&
{}-G'_T\Sslash_{hT}\gamma_5
-G_s^\perp\frac{\kslash_T\gamma_5}{M_h}
-H_{1T}\frac{i\sigma_{\mu\nu}\gamma_5S_{hT}^\mu P_h^\nu}{M_h}
-H_{1s}^\perp\frac{i\sigma_{\mu\nu}\gamma_5k_T^\mu P_h^\nu}{M_h^2}
+H_1^\perp\frac{\sigma_{\mu\nu}k_T^\mu P_h^\nu}{M_h^2}   
\nnn &&
{}-H_T^\perp\frac{i\sigma_{\mu\nu}\gamma_5S_{hT}^\mu k_T^\nu}{M_h}
+H\sigma_{\mu\nu}n_-^\mu n_+^\nu                         
-H_si\sigma_{\mu\nu}\gamma_5n_-^\mu n_+^\nu
\Bigg\}.
\end{eqnarray}
The generalized FF's depend on the longitudinal momentum
fraction $z$ and on the transverse momentum $\tv{k}$ and consist of
(integrated) linear combinations of the coefficient functions $B_i$. Since,
there are more FF's --- 21 up to ${\cal O}(1/Q)$ --- than the 12
independent coefficient functions $B_i$, a number of relations is
expected. Indeed, exploiting the connection between the FF's and the $B_i$ 
leads to nine relations. Focussing here on the seven 
{\em T-odd} functions $D_{1T}^\perp$, $H_1^\perp$, $D_L^\perp$,
$D_T$, $E_L$, $E_T$ and $H$ there are four relations between them
\begin{equation} 
\label{trivrel}
D_L^\perp(z,-z\tv{k})=-D_{1T}^\perp(z,-z\tv{k})
\end{equation} 
\begin{equation} 
\label{Drel}
D_T(z)=z^3\frac{d}{dz}\left[\frac{D_{1T}^{\perp(1)}(z)}{z}\right]
\;\Longrightarrow\;
\int_0^1\!dz\!
\left[\frac{D_T(z)}{z}+2D_{1T}^{\perp(1)}(z)\right]=0
\end{equation} 
\begin{equation} 
\label{Hrel}
H(z)=z^3\frac{d}{dz}\left[\frac{H_{1}^{\perp(1)}(z)}{z}\right]
\;\Longrightarrow\;
\int_0^1\!dz\!\left[\frac{H(z)}{z}+2H_1^{\perp(1)}(z)\right]=0
\end{equation}
\begin{equation} 
\label{Erel}
E_L(z)=z^3\frac{d}{dz}\left[\frac{E_{T}^{(1)}(z)}{z}\right]
\;\Longrightarrow\;
\int_0^1\!dz\!\left[\frac{E_L(z)}{z}+2E_T^{(1)}(z)\right]=0
\end{equation} 
Here, the upper index $(1)$ denotes the first moment in $\tv{k}$. 
The relations follow directly from the hermiticity properties of the fields
and parity reversal invariance; the RHS's of (\ref{Drel}), (\ref{Hrel}), and 
(\ref{Erel})
assume not too singular end-point behaviour.\par
One particularly important example for the occurence of {\em T-odd} 
FF's is the differential cross section for unpolarized electrons
scattering off a transversely polarized target provided the transverse
momentum of the produced hadron is measured [3,4] 
. 
\begin{eqnarray} 
\label{diffXsect-OT}
\lefteqn{
\frac{d\sigma_{OT}}{dx\;dy\;dz\;d^2P_{h\perp}}=
\frac{4\pi\alpha\;s}{Q^4} |{\bf S}_\perp| \sum_{a,\bar a} e_a^2
\;(1-y)\Bigg\{\sin(\phi_h+\phi_s)\;
I\!\left[\frac
{{\bf\hat h}\cdot \tv{k}}{M_h}\;x\,h_1^a\,H_1^{\perp a} \right]}
\\ &&\hspace{-9mm}
{}+\sin(3\phi_h-\phi_s)
I\!\left[\frac{\left(4({\bf\hat h}\cdot\tv{p})^2
               -\tv{p}^2\right){\bf\hat h}\cdot\tv{k}
               -2{\bf\hat h}\cdot\tv{p}\tv{p}\cdot\tv{k}}{M^2M_h^2}
\;x\,h_{1T}^{\perp a}\,H_1^{\perp a}\right]
\!\Bigg\}
+{\cal O}\left(\frac{1}{Q}\right)
\nonumber
\end{eqnarray}
with $I \left[({\bf\hat h}\cdot\tv{p})fD\right]\equiv
\int d^2p_T \; d^2k_T \; \delta^2(\tv{p}+\tv{q}-\tv{k}) \;
({\bf\hat h}\cdot\tv{p})\; f(x,\tv{p})D(z,\tv{k})$.\\
Eq.(\ref{diffXsect-OT}) contains two sine functions depending on different
combinations of the azimuthal angles $\phi_s$ and $\phi_h$ of the target spin
vector and the momentum of the produced hadron, respectively. The asymmetry 
caused by the first term is known as the `Collins effect' [3]. Both azimuthal 
asymmetries are of leading order in an expansion in
powers of $1/Q$. Leading order asymmetries involving {\em T-odd}
FF's are possible only in experimental quantities sensitive to transverse 
momentum (for a more detailed discussion on the role of transverse momentum in 
deep-inelastic processes see [5]).\par
Another example of observables involving {\em time
reversal odd} fragmentation functions is the number asymmetry of produced
hadrons in the scattering of unpolarized leptons off a transversely polarized
target.
\begin{equation}
\label{numasy}
\frac{{\textstyle
      \sigma(\ell\stackrel{\rightarrow}{H})
      -\sigma(\ell\stackrel{\leftarrow}{H})}}
     {{\textstyle
      \sigma(\ell\stackrel{\rightarrow}{H})
      +\sigma(\ell\stackrel{\leftarrow}{H})}}=
\frac{2(2-y)\sqrt{1-y}}{1-y+y^2/2}\;\frac{M_h}{Q}\;
\frac{\sum_{a,\bar a}e_a^2 h_1^a(x)
         \left(H^a(z)+2zH_1^{\perp(1)a}(z)\right)}
     {\sum_{a,\bar a}e_a^2 f_1^a(x)D_1^a(z)}
\end{equation}
Using Eq.(\ref{Hrel}) the occuring linear combination of {\em T-odd} FF's,
$H(z)+2zH_1^{\perp(1)}(z)$ can be simplified to the form 
$z\frac{d}{dz}\left[z\,H_1^{\perp(1)}(z)\right]$. This is a
typical example for the usefulness of the relations (\ref{trivrel}), 
(\ref{Drel}),(\ref{Hrel}) and (\ref{Erel}). \\[1mm]
{\footnotesize This work was supported by the foundation for Fundamental
Research on Matter (FOM) and the Dutch Organization for Scientific Research
(NWO).}
%
\vfill
{\small\begin{description}
\addtolength{\itemsep}{-2.3mm}
\item[1]
H.D.~Politzer, Nucl.~Phys.~B172, 349 (1980);\\ \hspace*{-6.4mm}
R.K.~Ellis, W.~Furmanski, R.~Petronzio, Nucl.~Phys.~B212, 29 (1983).
\item[2]
J.P.~Ralston, D.E.~Soper, Nucl.~Phys.~B152, 109 (1979);\\ \hspace*{-6.4mm}
R.D.~Tangerman, P.J.~Mulders, Phys.~Rev.~D51, 3357 (1995).
\item[3]
J.C.~Collins, Nucl.~Phys.~B396, 161 (1993).
\item[4]
P.J.~Mulders, R.D.~Tangerman, Phys.~Lett.~B352, 129 (1995);
and NPB 461, 197 (1996).
\item[5] 
R.~Jakob, A.~Kotzinian, P.J.~Mulders, J.~Rodrigues, these proceedings. 
\end{description}}

\end{document}